# Highly Stable and Reproducible Au Nanorod Arrays for Near-Infrared Optofluidic SERS sensor

*Samir Kumar, Takao Fukuoka\*, Kyoko Namura, and Motofumi Suzuki\**

Department of Micro Engineering, Graduate School of Engineering, Kyoto University, Katsura, Nishikyo, Kyoto 615-8540 Japan.

**KEYWORDS:** Au Nanorods, SERS, Raman Scattering, Optofluidics, Physical vapor deposition.

**ABSTRACT:** Surface-enhanced Raman spectroscopy (SERS) is a sensitive vibrational spectroscopy technique that can enable fast and non-destructive detection of trace molecules. SERS substrates are critical for the advancement of the SERS application. By incorporating SERS substrates into microfluidic devices, the function of microfluidic devices can be extended, and an efficient on-site trace analysis platform with powerful sensing capabilities can be realized. In this paper, we report the fabrication of a rapid and sensitive optofluidic SERS device using a unique Au nanorod array (AuNRA) with a plasmon resonance frequency in the near IR region. The highly stable and reproducible AuNRA were fabricated by a facile dynamic oblique angle deposition technique. A typical spectrum of 4,4-bipyridine (BPY) with enhanced peaks was observed within a few seconds after the injection of an aqueous solution BPY. Time-course measurements revealed an outstandingly quick response of SERS in this system. Using the AuNR microfluidic device, approximately $2\times10^{-12}$ mole molecules were enough to produce detectable SERS signals. This work demonstrates rapid and sensitive chemical sensing using an optofluidic device equipped with a unique noble metal nanorod array.

**INTRODUCTION**

Lab-on-chip devices are progressing towards integrating various unit operations, including nanostructure, surface modification, biological and/or chemical reactions, micro-channel design, and manufacturing. Advances in nanotechnology have accelerated the synergy combination of microfluidics and nanophotonics with optofluidics establishing itself in the fields of chemical and biological sensing.[1–3] Surface-enhanced Raman scattering (SERS) has provided the opportunity to design an ultrasensitive label-free lab-on-chip system owing to the high-performance assay with excellent sensitivity and selectivity of low sample volume.[4] Furthermore, due to the versatility of SERS as a non-destructive approach to chemical analysis and fingerprint characteristics, a wide range of SERS applications have been demonstrated in medicine, biochemistry, electrochemistry, and photocatalysis.[5–10]

Well-designed noble metal nanostructure employs the attractive nanophotonic phenomenon for SERS sensing. [11–15] Nevertheless, it is still challenging to develop facile methods to integrate SERS with highly sensitive, stable, and reproducible chip-biosensing systems capable of easy sample preparation and dynamic analysis of target molecules. Most researchers used colloidal aggregation of noble metal nanoparticles mixed with an analyte molecular solution to perform SERS measurements. However, the interaction of analyte molecules with the colloidal nanoparticles inside a conventional microfluidic channel is governed by the slow diffusion forces, and this leads to high signal variations owing to the instability of salt-induced colloidal aggregation and thus low reproducibility. [16–21] On the other hand, the solid-state substrates can generate more reproducible signals, but the fabrication technology employed (electron beam lithography and focused ion beam) to fabricate ordered and uniform nanostructures requires expensive equipment, cumbersome preparation process, and have difficulty in mass production for commercial use.

In this paper, we report a highly stable and reproducible Au nanorod arrays (AuNRAs) SERS platform (in near-infrared (IR) region) based optofluidic device fabricated by a facile dynamic oblique angle deposition (OAD) technique. For the application to the biological materials, the excitation in the near IR region is appropriate because the near IR is compatible with a biological tissues' transparency window for non-invasive examination of biological processes and pathogenic bacteria in a low fluorescence background.[22] We demonstrate the convenient and real-time SERS sensing capability of this AuNRAs optofluidics SERS platform through using an aqueous solution of Raman probe 4,4'-bipyridine (BPY) with concentrations ranging from 1 mM to 100 nM. These AuNRAs were highly reproducible and stable even after 90 days of fabrication.

## MATERIALS AND METHODS

### Fabrication of AuNRAs

The elongated AuNRAs were fabricated using a dynamic OAD technique, in which the deposition angle and/or in-plane direction of the substrate can be changed during deposition, thus allowing to control the morphology.[23–25] The detailed fabrication process of AuNRAs can be found elsewhere.[11,26] $SiO_2$ nano-columns with anisotropic surface morphology were deposited using the serial bideposition (SBD) method.[10,27] The vapor deposition angle $\alpha$ for $SiO_2$ was fixed at 78.6°, and the in-plane angle was swiftly rotated by 180° after each fixed amount of vapor deposition. By repeating 60 cycles of SBD SCL, a thickness of $d_{SiO2}$ = 600 nm was prepared, see Figure 1(b). Nanorod like anisotropic assemblies of Au nanoparticles was deposited on the top of the in-line $SiO_2$ nano-columns. Au was deposited up to thickness $t_{Au}$ of 10 nm at a deposition angle $\alpha_{Au}$ of 73.4°. The in-plane angle was not changed during the deposition of Au. The amount of deposition of Au was small enough to maintain the nano rod-like anisotropic ensemble of Au nanoparticles.

**Fabrication of Optofluidic platform**

Figure 1a shows a photograph of the polydimethylsiloxane (PDMS) AuNRAs optofluidic SERS device. 500 µm width ditches were prepared on the PDMS sheet by nanolithography. A 6 mm through hole was created in the middle of the ditch. A glass slide was placed on top of the ditch to make the microchannel, and AuNRAs chip (3 x 4 mm) was stuck on the other side of the PDMS sheet so as to close the through-hole (see Figure 1b). The volume of the sample solution required to fill up the sensing chamber was approximately 20 µl. Figure 1c presents the AuNRAs immersed in the injected solution through the microchannel.

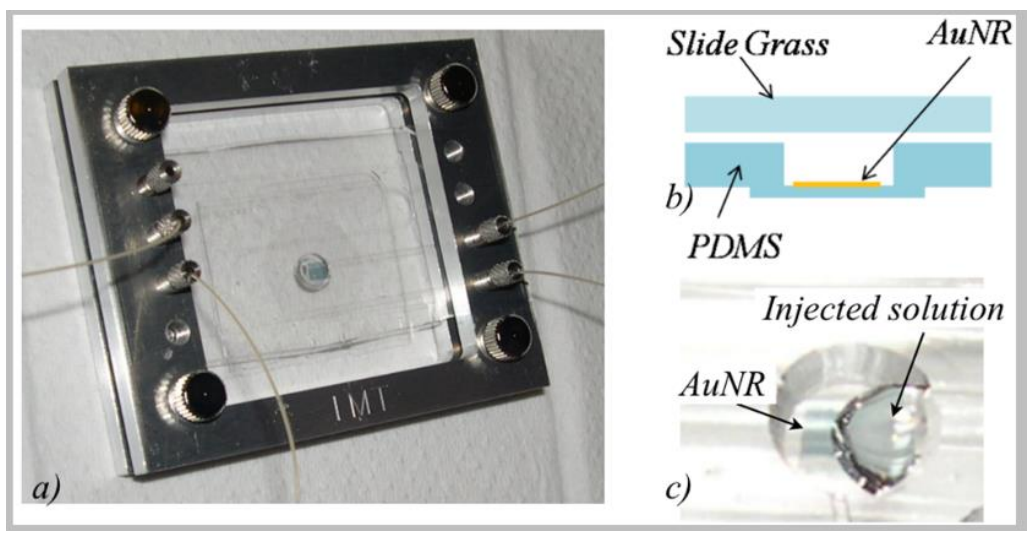

**Figure 1.** (a) A photograph of the polydimethylsiloxane (PDMS) AuNRAs optofluidic SERS device. (b) Cross section scheme of the sensing chamber. (c) Immersed AuNRAs with injected solution through the microchannel.

**SERS measurements**

In situ SERS measurements were performed after the injection of an aqueous solution of BPY to AuNRAs microchannel. Spectra were recorded using 785 nm laser (100 mW) Raman Spectrometer RAM-100S (Lambda-Vision Inc.).

**RESULTS AND DISCUSSIONS**

**Figure 2**a presents the bird's-eye view of the AuNRA morphology. The orange arrow represents the direction of Au vapor deposition, whereas the white arrows represent the direction of $SiO_2$ deposition. Figure 2b presents the anisotropic morphologies of the $SiO_2$ layer. The cross-section parallel to the $SiO_2$ flux incident plane reveals a narrow and long columnar structure. The anisotropic $SiO_2$ layer was self-assembled without using any lithography technique. When a few nm of Au is deposited at a large glancing angle on the anisotropic $SiO_2$ layer, Au sticks to the elevated part of the $SiO_2$ surface. Consequently, elongated semi-parallel Au nanoparticles representing the anisotropic surface morphology of the $SiO_2$ layer are obtained (see Figure 2c). Previously, we have demonstrated that when a small amount of Au is obliquely deposited, the morphology of AuNRAs depends on the growth parameters of the dynamic oblique

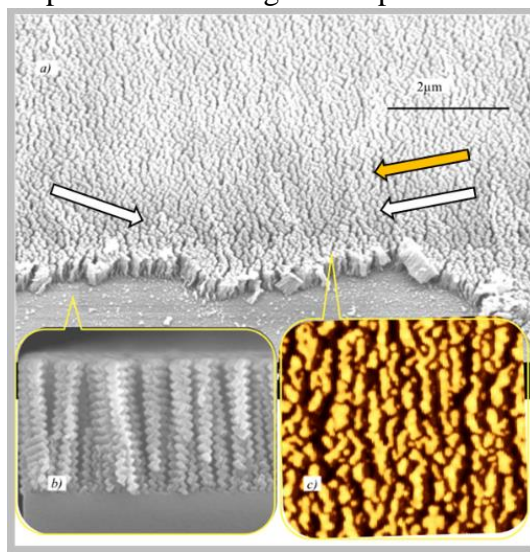

**Figure 2.** (a) SEM images of AuNRAs; (b) side view of in-line silica columns; and (c) closeup top view of nanorod-like anisotropic assembles of gold nanoparticles on the columns.

deposition.[11,23] In this study, we selected AuNRAs with plasmon resonance frequency, hence SERS activity, in the near IR region. The AuNRAs had a long-range uniformity $> 50 \times 50$ mm$^2$ with a density of AuNPs on SiO$_2$ columns of approximately 10-100 particles/μm$^2$.

Figure 3a presents the real-time SERS spectra of 5 μM aqueous solution of BPY with one second irradiation of 785 nm laser after every five seconds. The BPY solution was injected into the device after the passage of 40 s. Immediately after injection, the sample solution entered the sensing chamber with the SERS active AuNRAs, as shown in Figure 1c. Within 5 seconds, a typical spectrum of BPY with distinct peaks at 1000, 1200, 1265, and 1580 cm$^{-1}$ were observed, which are attributed to the pyridine ring breathing, ring deformation, C = C in-plane ring mode, and C = C stretching mode, respectively. [28,29] Time-course measurements revealed an outstandingly quick response of SERS in our system. While conventional colloidal methods intrinsically required salt-induced aggregation procedure, in contrast, our device with AuNRAs was able to apply in situ

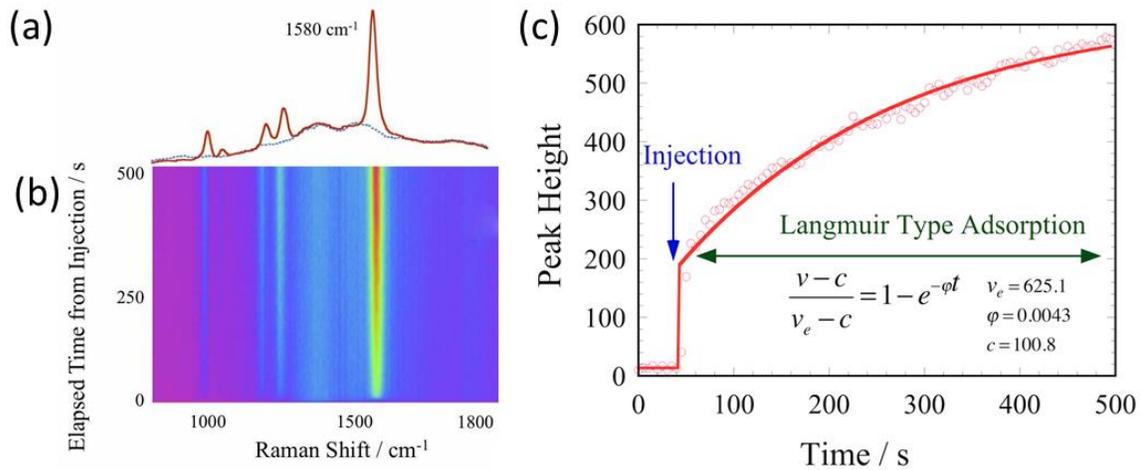

**Figure 3.** (a) Real time SERS spectra after injection of 5 μM BPY into the AuNRAs optofluidic device. Solid red line is the spectrum of BPY. Dotted line is the blank spectrum of AuNRAs without BPY. (b) Time course map of the BPY peaks intensities depending on the elapsed time after the BPY injection. (c) Time evolution of 1575 cm$^{-1}$ peak of BPY.

SERS measurements without aggregation pretreatment. Figure 3(c) shows that the typical time course of SERS intensity follows the Langmuir's equation of adsorption rate, which we believe, is one aspect of SERS adsorption phenomena. In this equation, ν represents adsorbed site of surface, in this case, it may be the hot spots. φ represents a kind of barrier for molecule moving onto the surface, one aspect is, for instance, contaminant on the surface.

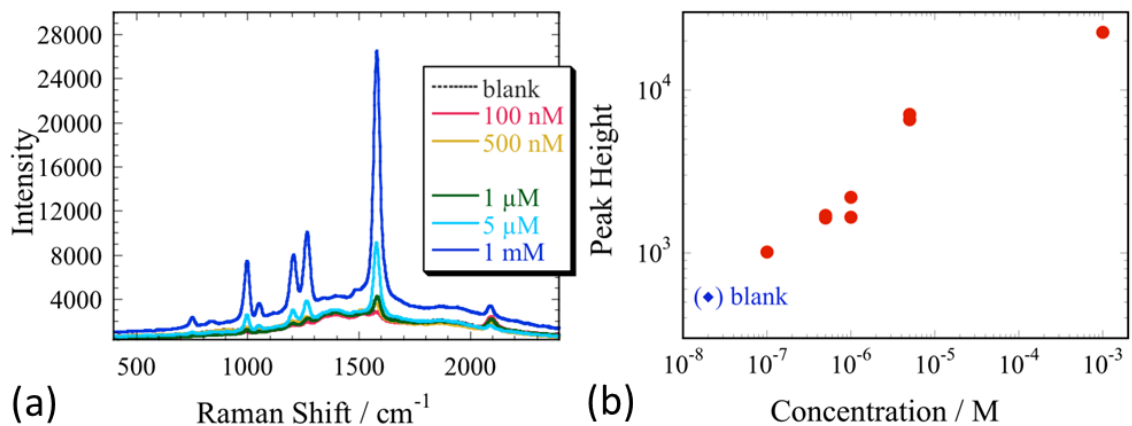

**Figure 4.** (a) SERS spectra of aqueous solution of BPY in the range of 100 nM–1 mM; (b SERS intensity of the most intense peak at 1575 cm$^{-1}$ as a function of BPY concentration.

To determine the sensitivity of the optofluidic SERS chip, SERS spectra for different BPY concentrations were acquired by 10-second irradiation, starting with the lowest BPY concentration. To avoid erroneous results due to contamination of the channels with residual molecules, after each measurement, the solution was suctioned out, and the microfluidic channel was flushed with water before higher concentration solutions were added and tested. Figure 4(a) shows the SERS spectra of an aqueous solution of BPY in the range of 100 nM–1 mM. Figure 4(b) shows the intensity of the most intense peak at 1575 cm$^{-1}$ as a function of BPY concentration. Taking the detection volume into account, almost $2\times10^{-12}$ moles molecules were enough to produce a significant detectable SERS signal. It should be noted that the normal Raman spectrum of 100 mM BPY was hardly observed without SERS active AuNRAs in our Raman spectrometer. Therefore, the apparent enhancement is estimated to be at least $10^6$.

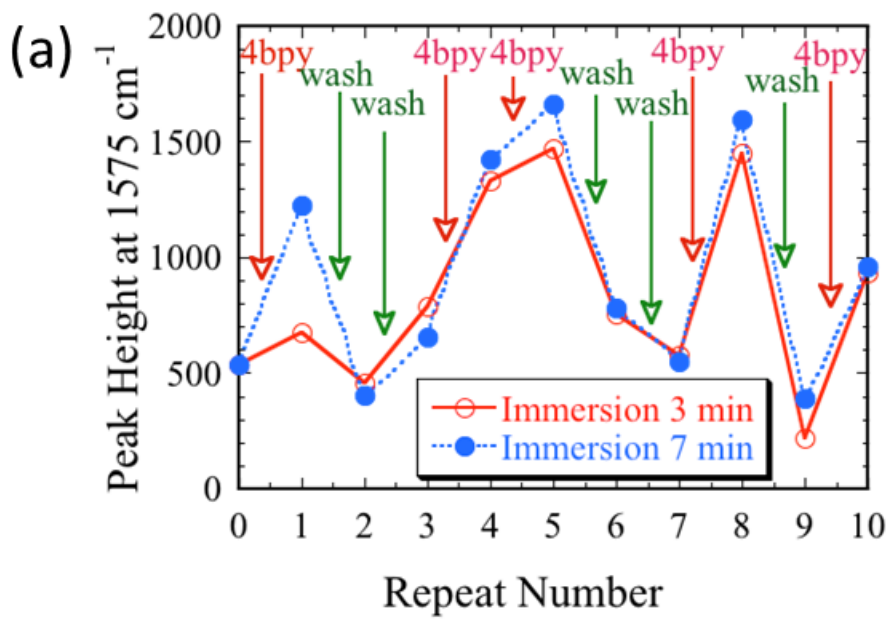

**Figure 5.** The variation is Raman intensity of 1575 cm$^{-1}$ peak after each injection and washing cycle.

The reusability of this AuNRAs optofluidic device is demonstrated through a series of repeated detection and washing experiments. Figure 5 presents the SERS spectra from 500 µM of an aqueous solution of BPY recorded before and after each immersion and washing step. An aqueous solution of BPY was injected and immersed for 3- (red curve) and 5-min (blue curve). It was found that the AuNR microfluidic device can be regenerated by flushing the system with deionized water between measurements without compromising the reusability of the system.

In addition to sensitivity, reproducibility, and stability over a long storage period are major concerns for any SERS substrates. Hence, SERS measurements of BPY with different concentrations were performed to test the stability and reproducibility of the AuNRAs. Figure 6(a) and 6(b) shows the SERS spectra of BPY solution two samples prepared under the same condition after 90 days, using a piece cut from each sample stored in a standard desiccator. The SERS response of AuNRAs declined slightly after 90 days as compared to that of fresh AuNRAs however, the SERS enhancement even after 90 days was high enough to use it as an efficient SERS sensor. We were able to detect BPY down to a concentration of $15.6 \times 10^{-9}$ g/mL. Since Au is chemically stable, we believe that the SERS properties should exhibit long-term stability, so the decrease in the SERS intensity may be attributed to the absorption of contamination on the Au nanorods.

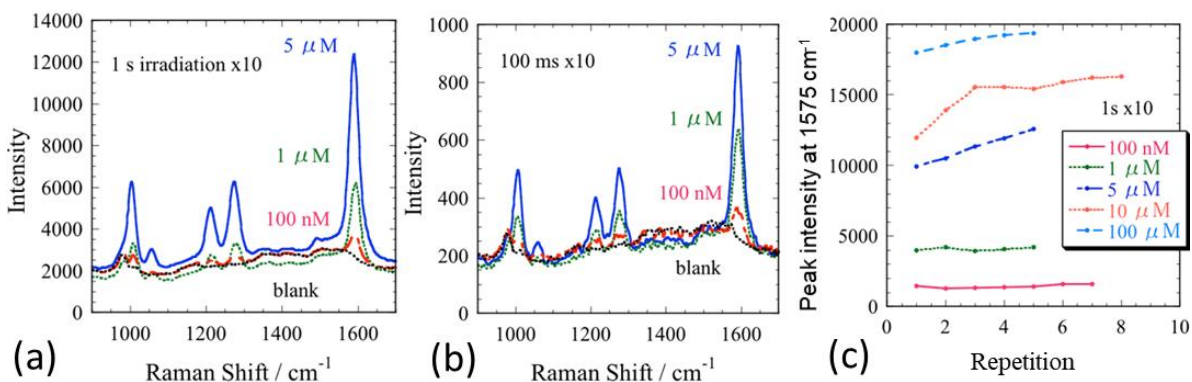

**Figure 6.** (a) SERS spectra of BPY solution on (a) fresh AuNRAs and (b) after 90 days, on samples prepared under the same condition. (c) Intensity of the 1575 cm$^{-1}$ peak for various concentrations of aqueous solution of BPY from different repetitions.

To assess the reproducibility of the AuNRAs, the intensity of the 1575 cm$^{-1}$ peak for 5 μM BPY solution from 7 different sites on the AuNRA substrate is shown in Figure 6(c). The variation of intensity was found to be within 5% for concentrations below 5 μM. However, for concentrations ≥ 5 μM, the variation in intensity was around 20% and was found to increase with the repeated measurements. The increase in the SERS intensity for the high concentrations of BPY can be explained, considering the photothermal effect. [30] AuNRAs can act as a nanoplasmonic antenna on the exposure of laser in the near IR region. This thermoplasmonic heating of the AuNRAs can produce a thermal gradient in the BPY solution near the metal nanoparticles. Consequently, the thermal gradient can cause a significant change in the SERS intensity by forcing the analyte molecules to form a circular vortex ring due to thermal diffusion and convection.[30] The SERS intensity changes by enhancement-reduction of photothermal gradients depending upon the position. However, the number of molecules may not be sufficient for the enrichment with the thermal gradient effect for the lower concentrations.

## CONCLUSION

In conclusion, we demonstrated an optofluidic device based on highly stable and reproducible AuNRAs. The rapid and sensitive analysis was sufficiently carried with a detection limit of BPY down to a concentration of $15.6 \times 10^{-9}$ g/mL using the AuNRAs on the microfluidic device. AuNRAs, we found to be an excellent and convenient nanostructure for SERS detection in optofluidics without any aggregation pretreatment. The combination of SERS and the microfluidic system can significantly contribute to broadening their application field.

## ASSOCIATED CONTENT

**Supplementary Information**.

## AUTHOR INFORMATION

**Corresponding Author**

*TF: tak_f@mpe.me.kyoto-u.ac.jp, MS: m-snki@me.kyoto-u.ac.jp

**Funding**

This work was supported by JST COI under Grant Number JPMJCE1307.

## ACKNOWLEDGMENT

We thank Dr. Kosuke Ishikawa of Kyoto University for assisting us with the SEM observations.

## REFERENCES

[1] D. Psaltis, S.R. Quake, C. Yang, Developing optofluidic technology through the fusion of microfluidics and optics, Nature. 442 (2006) 381–386. https://doi.org/10.1038/nature05060.

[2] X. Fan, I.M. White, Optofluidic microsystems for chemical and biological analysis, Nat. Photonics. 5 (2011) 591–597. https://doi.org/10.1038/nphoton.2011.206.

[3] S.H. Yazdi, I.M. White, Optofluidic Surface Enhanced Raman Spectroscopy Microsystem for Sensitive and Repeatable On-Site Detection of Chemical Contaminants, Anal. Chem. 84 (2012) 7992–7998. https://doi.org/10.1021/ac301747b.

[4] L. Chen, J. Choo, Recent advances in surface-enhanced Raman scattering detection technology for microfluidic chips, ELECTROPHORESIS. 29 (2008) 1815–1828. https://doi.org/10.1002/elps.200700554.

[5] S.S. Sinha, S. Jones, A. Pramanik, P.C. Ray, Nanoarchitecture Based SERS for Biomolecular Fingerprinting and Label-Free Disease Markers Diagnosis, Acc. Chem. Res. 49 (2016) 2725–2735. https://doi.org/10.1021/acs.accounts.6b00384.

[6] K. Xu, R. Zhou, K. Takei, M. Hong, Toward Flexible Surface-Enhanced Raman Scattering (SERS) Sensors for Point-of-Care Diagnostics, Adv. Sci. 6 (2019) 1900925. https://doi.org/10.1002/advs.201900925.

[7] X. Ma, Y. Xia, L. Ni, L. Song, Z. Wang, Preparation of gold nanoparticles–agarose gel composite and its application in SERS detection, Spectrochim. Acta. A. Mol. Biomol. Spectrosc. 121 (2014) 657–661. https://doi.org/10.1016/j.saa.2013.11.111.

[8] S. Kumar, D. K. Lodhi, P. Goel, Neeti, P. Mishra, J. P. Singh, A facile method for fabrication of buckled PDMS silver nanorod arrays as active 3D SERS cages for bacterial sensing, Chem. Commun. 51 (2015) 12411–12414. https://doi.org/10.1039/C5CC03604F.

[9] S. Kumar, D. K. Lodhi, J. P. Singh, Highly sensitive multifunctional recyclable Ag–TiO 2 nanorod SERS substrates for photocatalytic degradation and detection of dye molecules, RSC Adv. 6 (2016) 45120–45126. https://doi.org/10.1039/C6RA06163J.

[10] S. Kumar, Y. Doi, K. Namura, M. Suzuki, Plasmonic Nanoslit Arrays Fabricated by Serial Bideposition: Optical and Surface-Enhanced Raman Scattering Study, ACS Appl. Bio Mater. 3 (2020) 3226–3235. https://doi.org/10.1021/acsabm.0c00215.

[11] M. Suzuki, K. Nakajima, K. Kimura, T. Fukuoka, Y. Mori, Au Nanorod Arrays Tailored for Surface-Enhanced Raman Spectroscopy, Anal. Sci. 23 (2007) 829–833. https://doi.org/10.2116/analsci.23.829.

[12] M. Suzuki, W. Maekita, Y. Wada, K. Nakajima, K. Kimura, T. Fukuoka, Y. Mori, In-line aligned and bottom-up Ag nanorods for surface-enhanced Raman spectroscopy, Appl. Phys. Lett. 88 (2006) 203121. https://doi.org/10.1063/1.2205149.

[13] A. Rajput, S. Kumar, J.P. Singh, Vertically standing nanoporous Al–Ag zig-zag silver nanorod arrays for highly active SERS substrates, Analyst. 142 (2017) 3959–3966. https://doi.org/10.1039/C7AN00851A.

[14] V. Bochenkov, J. Baumberg, M. Noginov, F. Benz, H. Aldewachi, S. Schmid, V. Podolskiy, J. Aizpurua, K. Lin, T. Ebbesen, A.A. Kornyshev, J. Hutchison, K. Matczyszyn, S. Kumar, B. de Nijs, F.R. Fortuño, J.T. Hugall, P. de Roque, N. van Hulst, S. Kotni, O. Martin, F.J.G.

de Abajo, M. Flatté, A. Mount, M. Moskovits, P. Ginzburg, D. Zueco, A. Zayats, S.-H. Oh, Y. Chen, D. Richards, A. Belardini, P. Narang, Applications of plasmonics: general discussion, Faraday Discuss. 178 (2015) 435–466. https://doi.org/10.1039/C5FD90025E.

[15] R. Takahashi, T. Fukuoka, Y. Utsumi, A. Yamaguchi, Microfluidic devices with three-dimensional gold nanostructure for surface enhanced Raman scattering, in: 8th Annu. IEEE Int. Conf. NanoMicro Eng. Mol. Syst., IEEE, Suzhou, China, 2013: pp. 722–725. https://doi.org/10.1109/NEMS.2013.6559830.

[16] P. Measor, L. Seballos, D. Yin, J.Z. Zhang, E.J. Lunt, A.R. Hawkins, H. Schmidt, On-chip surface-enhanced Raman scattering detection using integrated liquid-core waveguides, Appl. Phys. Lett. 90 (2007) 211107. https://doi.org/10.1063/1.2742287.

[17] I.-H. Chou, M. Benford, H.T. Beier, G.L. Coté, M. Wang, N. Jing, J. Kameoka, T.A. Good, Nanofluidic Biosensing for β-Amyloid Detection Using Surface Enhanced Raman Spectroscopy, Nano Lett. 8 (2008) 1729–1735. https://doi.org/10.1021/nl0808132.

[18] D. Choi, T. Kang, H. Cho, Y. Choi, L.P. Lee, Additional amplifications of SERSvia an optofluidic CD-based platform, Lab Chip. 9 (2009) 239–243. https://doi.org/10.1039/B812067F.

[19] T. Fukuoka, A. Yamaguchi, R. Hara, T. Matsumoto, Y. Utsumi, Y. Mori, Application of gold nanoparticle self-assemblies to unclonable anti-counterfeiting technology, in: 2015 Int. Conf. Electron. Packag. IMAPS Asia Conf. ICEP-IAAC, IEEE, Kyoto, Japan, 2015: pp. 432–435. https://doi.org/10.1109/ICEP-IAAC.2015.7111051.

[20] C.-J. Heo, S.-H. Kim, S.G. Jang, S.Y. Lee, S.K. Kim, S.-M. Yang, Metal nanograil arrays with tunable multiple dipolar plasmon modes in integrated optofluidic devices for ultrasensitive sensing of biomolecules, in: E.A. Dobisz, L.A. Eldada (Eds.), San Diego, California, USA, 2008: p. 703915. https://doi.org/10.1117/12.794330.

[21] J. Guo, F. Zeng, J. Guo, X. Ma, Preparation and application of microfluidic SERS substrate: Challenges and future perspectives, J. Mater. Sci. Technol. 37 (2020) 96–103. https://doi.org/10.1016/j.jmst.2019.06.018.

[22] L. Chen, N. Mungroo, L. Daikuara, S. Neethirajan, Label-free NIR-SERS discrimination and detection of foodborne bacteria by in situ synthesis of Ag colloids, J. Nanobiotechnology. 13 (2015) 45. https://doi.org/10.1186/s12951-015-0106-4.

[23] M. Suzuki, W. Maekita, Y. Wada, K. Nagai, K. Nakajima, K. Kimura, T. Fukuoka, Y. Mori, Ag nanorod arrays tailored for surface-enhanced Raman imaging in the near-infrared region, Nanotechnology. 19 (2008) 265304. https://doi.org/10.1088/0957-4484/19/26/265304.

[24] S. Kumar, P. Goel, D.P. Singh, J.P. Singh, Highly sensitive superhydrophobic Ag nanorods array substrates for surface enhanced fluorescence studies, Appl. Phys. Lett. 104 (2014) 023107. https://doi.org/10.1063/1.4861836.

[25] S. Kumar, P. Goel, D.P. Singh, J.P. Singh, Fabrication of superhydrophobic silver nanorods array substrate using glancing angle deposition, in: AIP Conf. Proc., 2014: p. 872. https://doi.org/10.1063/1.4872786.

[26] M. Suzuki, K. Nakajima, K. Kimura, T. Fukuoka, Y. Mori, Physically Self-assembled Au Nanorod Arrays for SERS, MRS Proc. 951 (2006) 0951-E09-35. https://doi.org/10.1557/PROC-0951-E09-35.

[27] S. Li, M. Suzuki, K. Nakajima, K. Kimura, T. Fukuoka, Y. Mori, An approach to self-cleaning SERS sensors by arraying Au nanorods on TiO 2 layer, Nanocoatings. 6647 (2007) 66470J. https://doi.org/10.1117/12.731991.

[28] T. Lu, T.M. Cotton, R.L. Birke, J.R. Lombardi, Raman and surface-enhanced Raman spectroscopy of the three redox forms of 4,4'-bipyridine, Langmuir. 5 (1989) 406–414. https://doi.org/10.1021/la00086a021.

[29] M. Muniz-Miranda, B. Pergolese, A. Bigotto, A. Giusti, Stable and efficient silver substrates for SERS spectroscopy, J. Colloid Interface Sci. 314 (2007) 540–544. https://doi.org/10.1016/j.jcis.2007.05.089.

[30] T. Kang, S. Hong, Y. Choi, L.P. Lee, Plasmonics: The Effect of Thermal Gradients in SERS Spectroscopy (Small 23/2010), Small. 6 (2010) 2622–2622. https://doi.org/10.1002/smll.201090082.